# Fabrication of a high-resolution smartphone spectrometer for education using a 3D printer

Yura Woo and Young-Gu Ju

Department of Physics Education, Kyungpook National University, 80 Daehakro, Bukgu, Daegu, 41566, Korea

E-mail: ygju@knu.ac.kr

Abstract

In this paper, we present the details of the development of a smartphone spectrometer for education using a 3D printer and characterized the performance by comparison with a paper craft spectrometer. The optical design and the narrow slit used in the build resulted in the formation of accurate images of the slit on the image sensor leading to a superior resolution compared to the paper craft spectrometer. Increasing the exposure time of the phone's camera revealed the fine structure of a spectrum with high resolution. The baffle structure inside the spectrometer proved to be effective in removing noise when the exposure time was increased. We expect that the 3D printed smartphone spectrometer proposed in this paper can be useful as an education tool for students to understand the various aspects of light, atoms, chemistry, and physics.

1. Introduction

An optical spectrometer is an instrument used to analyse the consistent components of light. The separated components or spectrum provides information about the source or the material from which light is emitted, transmitted, reflected or scattered. The importance of spectral data is well-outlined in textbooks on physics and chemistry. In general, a basic course in modern physics usually includes a description of the Bohr model of the atom and its success in accounting for the emission spectrum of the hydrogen atom [1]. Since the spectrum provides information about the energy levels of an atom, it is often used to analyse the chemical components in astronomy, chemistry, biosensors and so on [2-5]. The splitting of light into different paths according to the wavelength also has many applications in the fabrication of devices used in optics and optical telecommunications [6]. The importance of a spectrometer in science results in part to its frequent appearance in science textbooks for high school students, although the principle of operation is typically not reviewed in detail.

Recently, a few internet sites have been created to provide detailed documentation on the construction of a smartphone spectrometer using paper and a compact disc as a "Do it yourself"(DIY) approach or outreach education scheme [7, 8]. Since the paper craft spectrometer consists of low-cost components, it is quite accessible to most young students and allows them to have fun while learning some basic science. The paper craft spectrometer exhibits quite good splitting of the light to some extent. However, when the spectrometer is exposed to a gas discharge lamp which is supposed to show a collection of distinct fine lines, the observed line width is significantly larger than those represented in standard textbooks. This can be disappointing from the viewpoint of students and educators, even though the performance limitation is readily attributable to the low-cost components. In fact, the limited performance seems to stem from the lack of an optical design rather than cost.

In this paper, we provide a detailed overview on the construction of a sophisticated smartphone spectrometer for educational purposes using a 3D printer and compare the performance to a paper craft spectrometer in terms of resolution and sensitivity. We anticipate that the improved smartphone spectrometer will be a valuable tool for students and educators as they explore advanced concepts in science.

2. Paper craft smartphone spectrometer

Prior to the investigation involving the 3D printed smartphone spectrometer, we built a paper craft spectrometer to evaluate its performance. The construction of the test model was performed according to the instructions on the spectrometer's website [6]. The paper craft spectrometer was made using a downloadable template which is printed on paper and folded along highlighted edges as shown in Fig. 1-(a). The assembled spectrometer had the reflection grating on the aperture as shown in Fig. 1-(b). The aperture of the spectrometer is in contact with the external optics of the smartphone camera. The grating was made by cutting a digital video disc (DVD) in order to lower the cost. One of the most important features of the paper craft spectrometer is the slit, which was made by cutting out a thin rectangular white region from the centre of the template as shown in Fig. 1-(a).

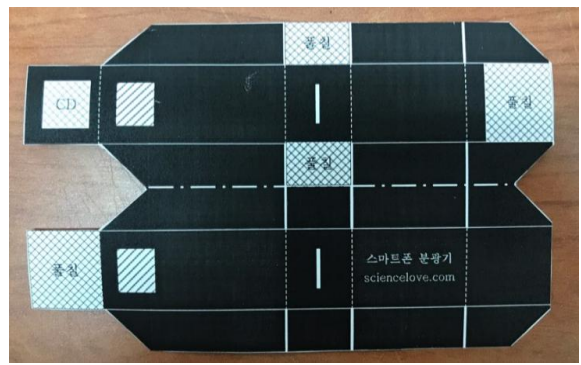

(a)

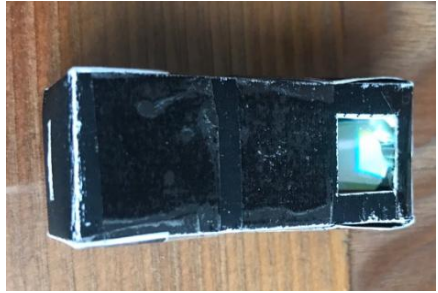

(b)

Fig. 1. Papercraft smartphone spectrometer: (a) Unfolded view before assembly (b) The assembled paper craft spectrometer shows the diffraction grating on the right and the slit on the left.

The spectrum from a mercury lamp acquired using the fabricated spectrometer is shown in Fig. 2. The images were taken at different exposure times. In this investigation, the smartphone used was the LG G4 which has a manual camera mode and can extend the exposure time up to 30 s. A long exposure time is a great benefit when the signal from the source is weak. The spectrum exhibits main spectrum lines which become more distinct as the exposure time is increased. However, the linewidth of the spectrum is much larger than of a conventional spectrometer used in the laboratory. These broad spectral lines are also observed in both reference websites of the papercraft spectrometer [7, 8]. This indicates that the lack of resolution is a common feature of the papercraft spectrometer. The low resolution is acceptable when the spectrometer is used for demonstrating basic principles of spectrometry. However, this limitation is a major obstacle in the case of quantitative analysis or when comparing experimental spectra with reference data.

There are two primary reasons for the low resolution of the papercraft spectrometer. One is the wide slit of the design, and the other is the distance between the slit and the smartphone camera. In fact, the spectrum lines are the images of the slit viewed through the grating and smartphone camera. In general, the slit width of the papercraft spectrometer is a few millimetres due to the limitation of paper cutting. A wider slit results in a wider image on the image sensor of the smartphone camera. Furthermore, the length of the papercraft spectrometer is approximately 50 mm, which is much smaller than the minimum object distance of the smartphone camera. The short distance between the slit and camera causes a defocusing error which blurs the slit image.

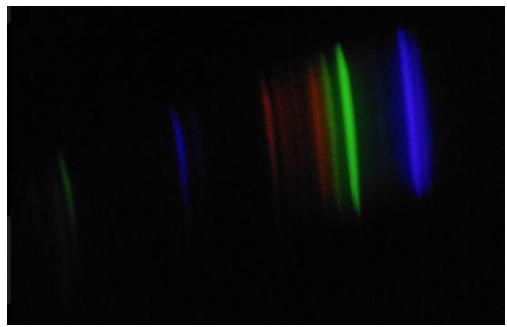

(a)

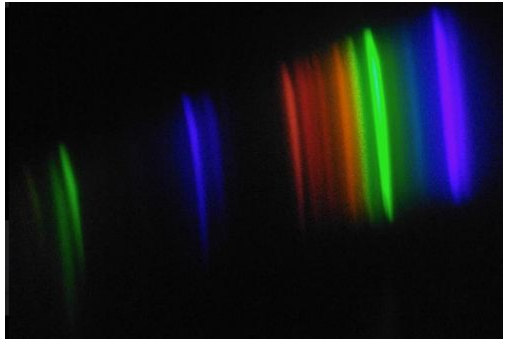

(b)

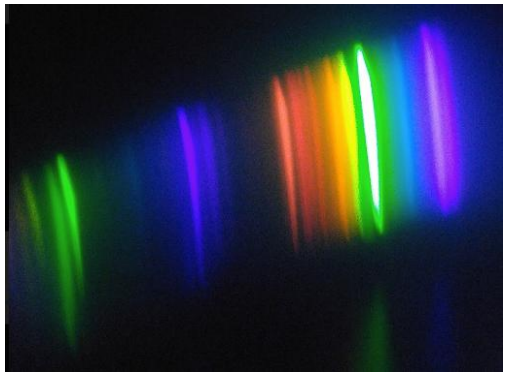

(c)

Fig. 2. Spectra of Hg lamp taken by the papercraft smartphone spectrometer with various exposure times: (a) 1/30 s (b) 1/8 s (c) 1/2 s

3. Fabrication of 3D printed smartphone spectrometer

In order to improve the performance of the smartphone spectrometer, we employed optical design principles in the development of a new spectrometer and used a 3D printer to fabricate the housing. Currently, 3D printers are inexpensive and readily available in most schools. The advantage of these instrument in the fabrication process is that they allow individuals to realize complicated structures without regard to fabrication skills, provided that they are familiar with computer-aided design (CAD) software, some of which is freely available online.

The assembly view of the designed spectrometer is shown in Fig. 3. The basic structure of the new designed is similar to that of the papercraft model. The major changes include the use of a separate slit and the modification of the length of the device to solve the aforementioned problems. The separate slit is used to implement a narrow slit which can be made on a glass plate using the doctor blade method [9]. This approach is one of the simplest methods to form a thin film on a substrate. In our case, silver paint was coated on a glass plate using the doctor blade method. After the film was dried, another blade was used to draw a line to form a slit on the silver film.

Images of the slit installed in the spectrometer were taken to measure the slit width as shown in Fig. 4. The slit width was determined to be approximately 30 μm using a calibration microscope. Since the length of the spectrometer is about 80 mm and the focal length of the smartphone camera is about 3 mm, the magnification factor is about 0.038. This means that the geometric image of the slit on the image sensor should be about 1.1 μm, which is about the pixel size of the image sensor. The length of the spectrometer was set to be about 80 mm, which is longer than the minimum object distance of the smartphone camera used in the experiment. The modifications to the slit and the device length are the major changes that were made to achieve an improved resolution of the spectrometer. We replaced the DVD disc with a relatively inexpensive reflection grating #43-751 purchased from Edmund optics to further improve the performance of the device. The grating has 1200 grooves and is 12.7 mm wide with a central wavelength of 500 nm. In addition, a baffle structure was installed near the slit and before the grating. A baffle structure is often used in telescope design to remove stray light which unexpectedly arrives at the image sensor. The fabricated baffle and its installation are shown in Fig. 5.

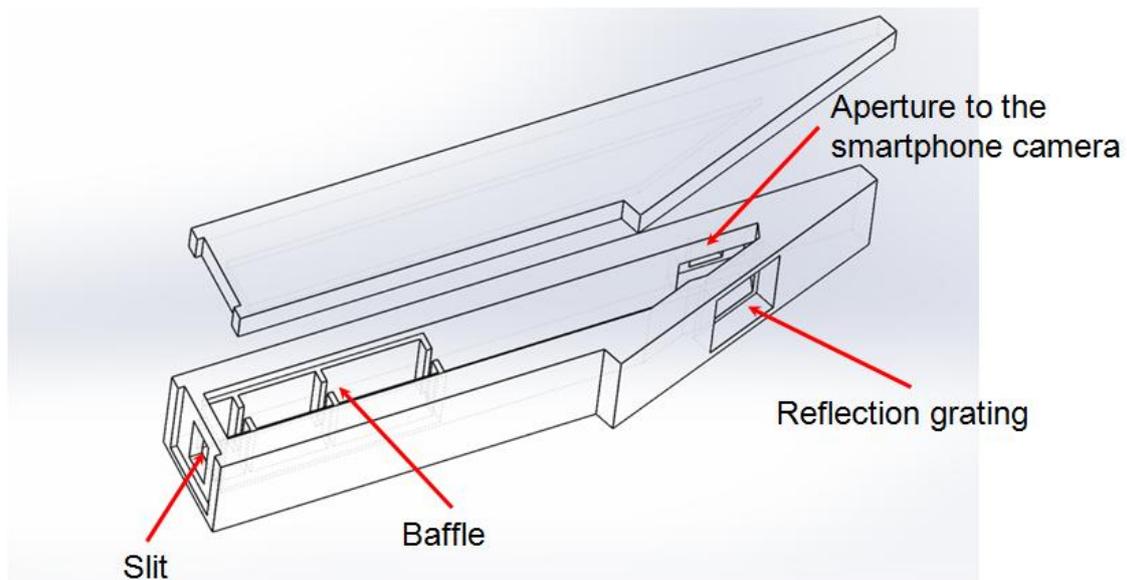

Fig. 3. Assembly view of the 3D printed smartphone spectrometer

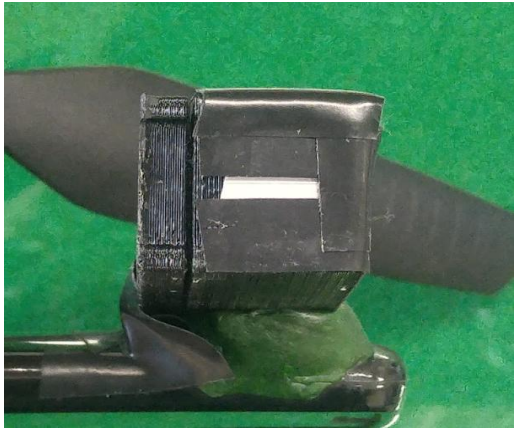

(a)

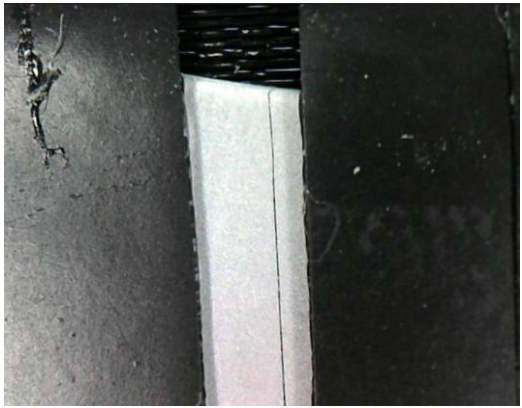

(b)

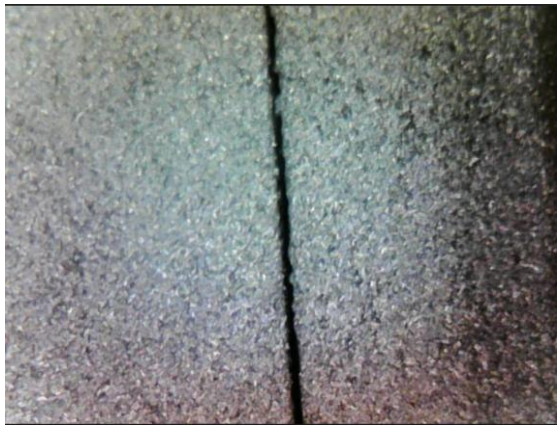

(c)

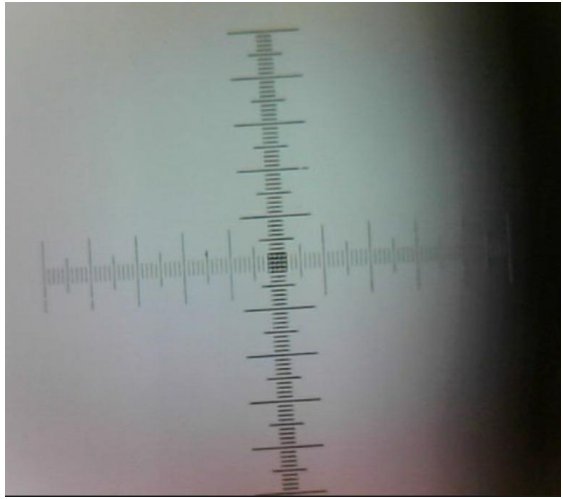

(d)

Fig. 4. The slit used in the 3D printed smartphone spectrometer: (a) low magnification (b) medium magnification (c) high magnification (d) the ruler at the high magnification

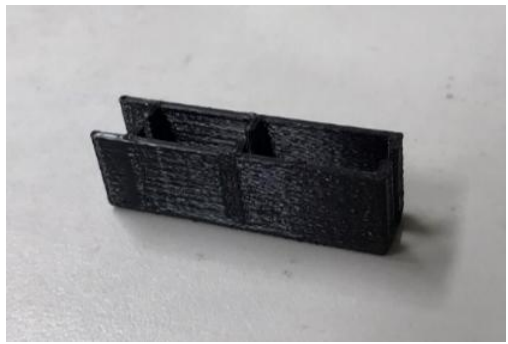

(a)

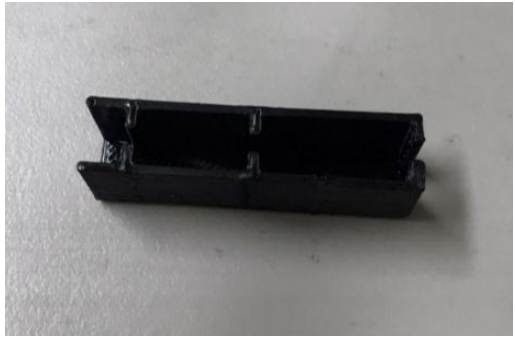

(b)

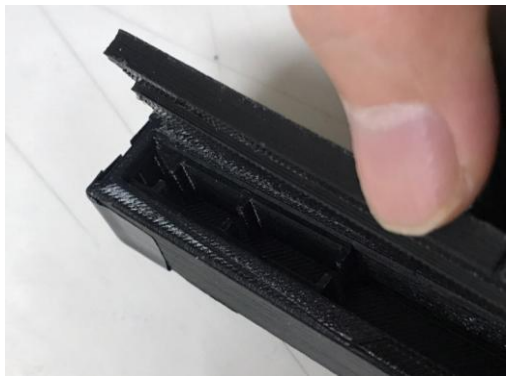

(c)

Fig. 5. The baffle structure used in the 3D printed smartphone spectrometer to avoid stray light: (a) oblique view (b) top view (c) the baffle installed inside the 3D smartphone spectrometer

4. Results and discussion

The fabricated smartphone spectrometer underwent performance testing by measuring the emission spectrum from several gas discharge lamps including Hg, Ar, and Ne. The spectrometer installed on the smartphone is shown in Fig. 6. A tripod was used to avoid movement of the camera during the exposure time. Mechanical stabilization is especially important when the exposure time is long for acquiring images of weak spectrum lines.

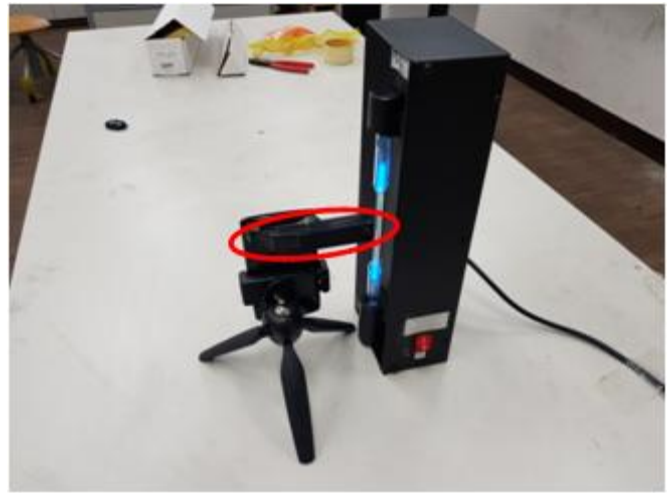

(a)

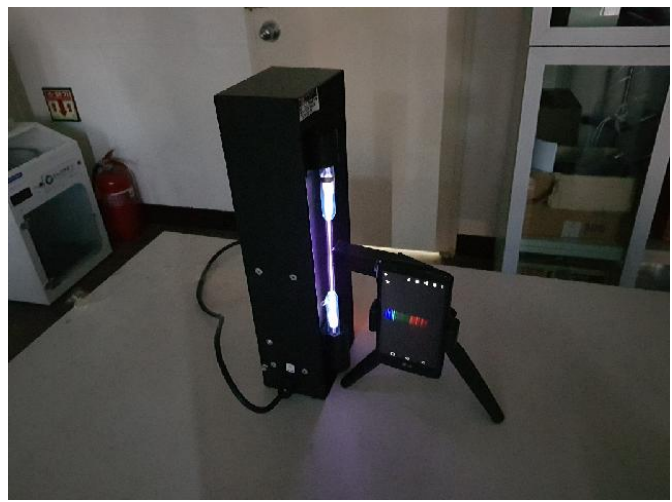

(b)

Fig. 6. 3D printed smartphone spectrometer with gas discharge lamp: (a) the red circle indicates the spectrometer (b) the front view showing the display.

The spectrum of the Hg lamp acquired by the 3D printed spectrometer is shown in Fig. 7. The improved resolution is evident when comparisons are made with the spectra from the papercraft spectrometer. As the exposure time is increased, fine structure starts to appear. Although the strong spectrum lines get broader, this is due to the saturation of the signal and not to the deterioration of the resolution. The linewidth of the finest spectral line is about 4 ~ 5 pixels wide. It is wider than the width expected from the image of the slit because the smartphone camera suffers from aberrations and diffraction. Therefore, the resolution of the spectrometer corresponds to the image resolution of the smartphone camera. We were able to obtain very fine resolution and high sensitivity from the newly fabricated smartphone spectrometer when compared to the papercraft spectrometer. In the case of the former, the observed spectrum of Ar and Ne lamps yielded results of similar quality which clearly depicted fine structures, as seen in Fig. 8.

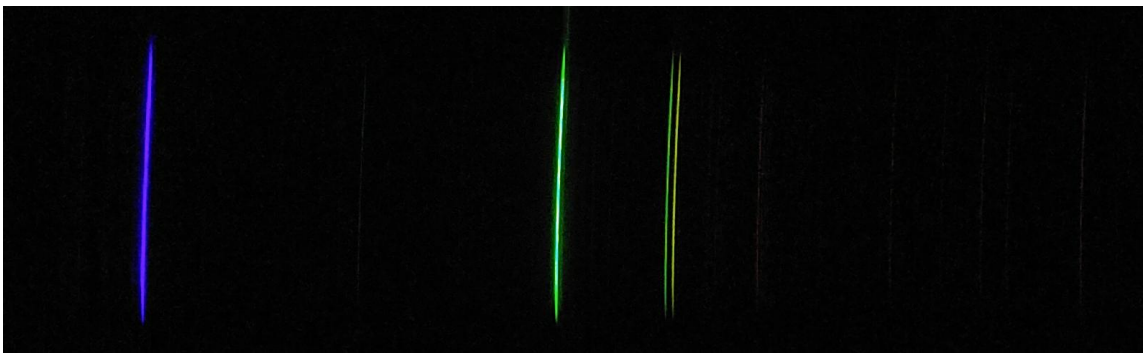

(a)

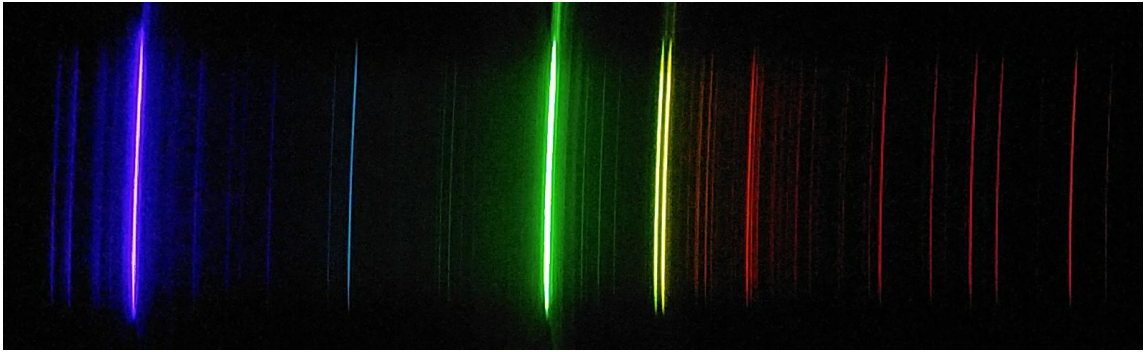
(b)

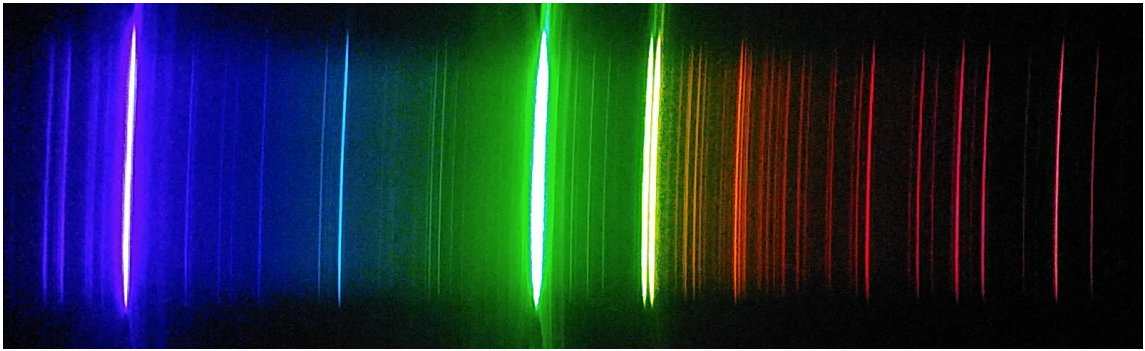
(c)

Fig. 7. Spectra of Hg lamp taken by the 3D printed smartphone spectrometer with the various exposure times: (a) 1/4 s (b) 2 s (c) 8 s

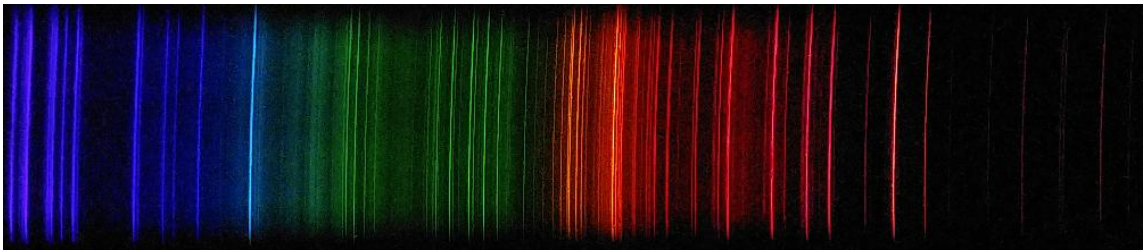
(a)

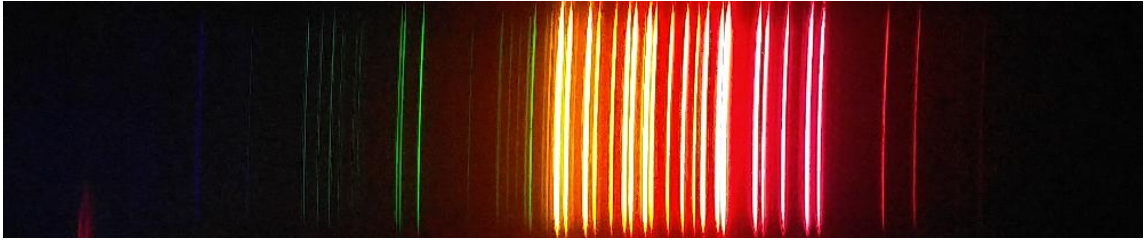

(b)

Fig. 8. Spectra of Ar and Ne lamp taken by the 3D printed smartphone spectrometer: (a) Ar spectrum with an exposure time of 8 s (g) Ne spectrum with an exposure time of 1/2 s

In order to evaluate the effect of the baffle, we took photos of the Hg spectrum with and without the baffle structure. The results are displayed in Fig. 9. When the baffle structure was absent, the spectrum suffered from wide spreading of the energy or large noise centring around the strong spectral line as seen in Fig. 9-(b), which is not seen in Fig. 9-(a).  This type of noise becomes more apparent as the exposure time increases.

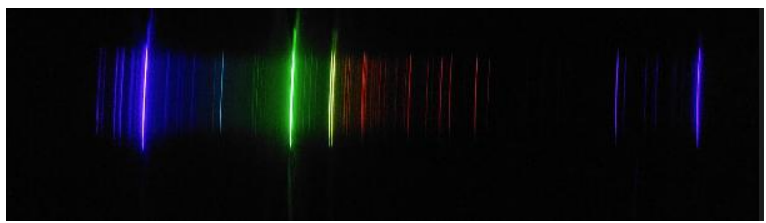

(a)

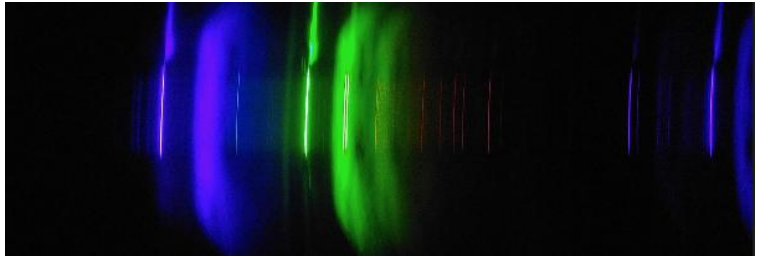

(b)

Fig 9. Spectra of mercury (a) with baffle and (b) without the baffle

The exposure time was 4 s.

The image of the spectrum can be converted into a quantitative plot as shown in Fig. 10. At first, the brightness information extracted from the picture using the ImageJ [10] software provides the spectral intensity as a function of pixel position. The wavelength of the strong spectral lines can be easily calibrated with a conventional spectrometer. The corresponding spectral lines in the image acquired by the 3D printed spectrometer can then be identified with their respective wavelengths. In the case of the Hg lamp used in the experiment, the wavelengths 446.80 nm, 540.25 nm, 582.18 nm, and 585.59 nm were used for calibration. In this manner, the pixel position was converted into a wavelength, and the intensity of the spectral lines can be plotted as a function of this parameter as seen in Fig. 10. The error identified in the plot shown in Fig. 10 is due to the truncation of the peaks which is likely the result of the saturation of strong signals due to the long exposure time. The truncation error of strong spectral lines is inevitable when weak signals in the spectrum are recording with long exposure times.

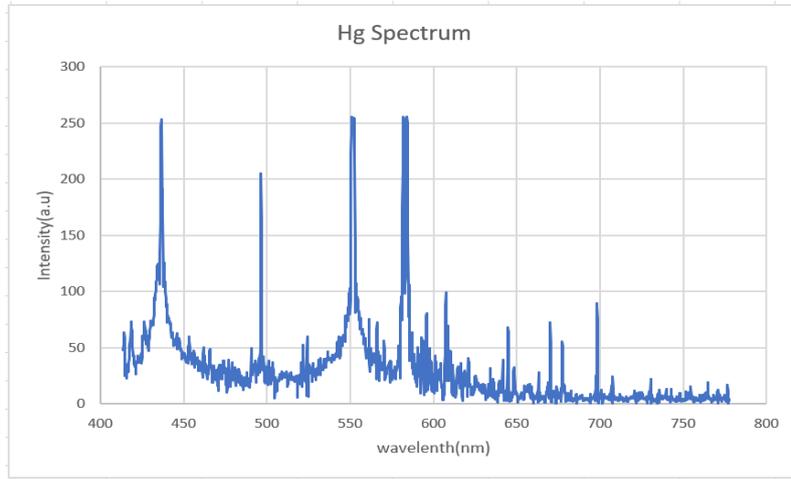

(a)

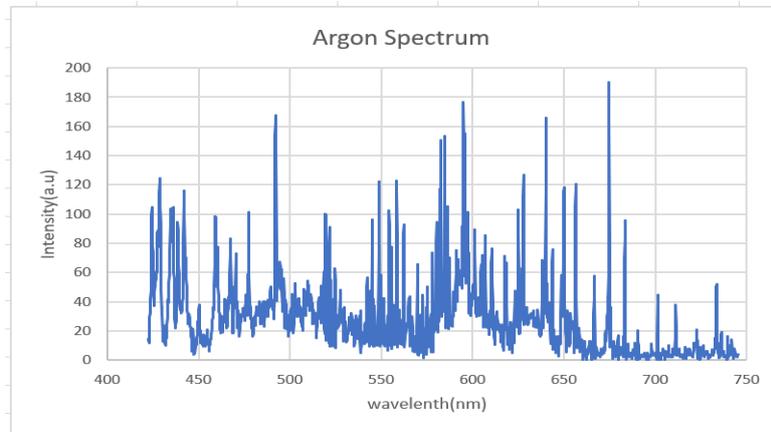

(b)

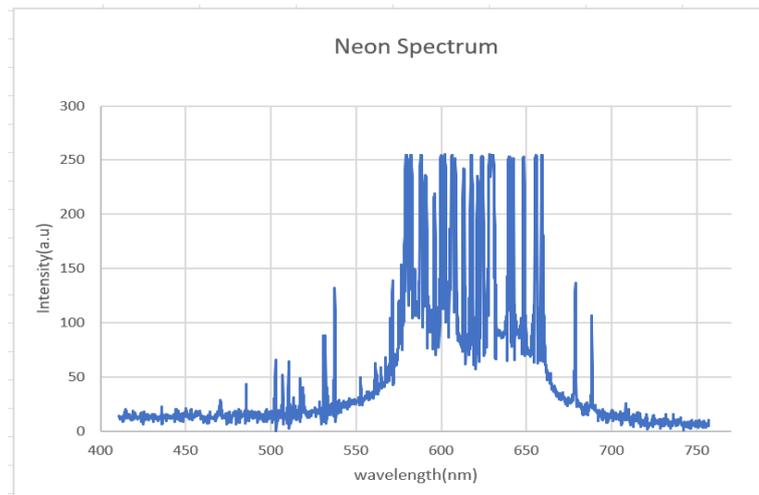

(c)

Fig. 10. Spectral intensity plots as a function of wavelength were extracted from the spectral images. (a) Hg   (b) Ar   (c) Ne

5. Conclusions

In this paper, we developed a smartphone spectrometer for educational purposes using a 3D printer and characterized the performance by comparing it with a papercraft spectrometer.

Firstly, we made a simple smart-phone spectroscope using a design template printed on paper which was folded along specified edges to create housing. A reflection grating made of a compact disc was inserted into the housing. The fabricated spectrometer revealed some strong spectral lines of several gas discharge lamps although the linewidths of the spectrum were large compared to that acquired using a conventional laboratory spectrometer. This was ascribed to the absence of appropriate optical design with respect to the resolution of the spectrometer. The images of the spectra were acquired using a smartphone in a professional setting, and the phone was fixed to a tripod stand. The exposure time of the phone camera was increased as required to amplify low-intensity signals in the spectrum.

Secondly, we fabricated a smartphone spectrometer using a 3D printer, a reflection grating, and a custom designed slit. The spectrometer was attached to a smartphone with a tripod stand. Since optical design principles and a narrow slit was used, an accurate image of the slit was formed on the image sensor leading to improved resolution compared to that of the papercraft spectrometer. Increasing the exposure time of the phone camera revealed the fine structure of the observed spectrum together with a high resolution. A baffle structure placed inside the spectrometer proved to be effective in the elimination of noise when the exposure time was increased.

The 3D printed smartphone spectrometer can facilitate the observation of atomic spectra with good resolution, which is typically a challenging task except with the use of expensive and sophisticated equipment. We expect that the 3D printed smartphone spectrometer proposed in this paper will be a useful educational tool for students, which will aid in their understanding of important aspects of light, atoms, chemistry, and physics.